\documentclass{article}

\usepackage{PRIMEarxiv}

\usepackage[utf8]{inputenc} 
\usepackage[T1]{fontenc}    
\usepackage{hyperref}       
\usepackage{url}            
\usepackage{booktabs}       
\usepackage{amsfonts}       
\usepackage{nicefrac}       
\usepackage{microtype}      
\usepackage{lipsum}
\usepackage{fancyhdr}       
\usepackage{graphicx}       
\graphicspath{{media/}}     
\usepackage{subcaption}
\usepackage[ruled,vlined]{algorithm2e}
\usepackage{amssymb, amsmath}
\usepackage[toc,page]{appendix}
\usepackage{balance}
\title{MAMRL: Exploiting  Multi-agent Meta Reinforcement Learning in WAN Traffic Engineering
\thanks{\textit{\underline{Citation}}: 
\textbf{Authors. Title. Pages.... DOI:000000/11111.}} 
}

\author{
 Shan Sun \\
  University of California, Riverside \\
   \texttt{ssun029@ucr.edu} \\
  \And
  Mariam Kiran\\
  Lawrence Berkeley National Laboratory\\
  \texttt{mkiran@es.net} \\
   \And
  Wei Ren\\
University of California, Riverside\\
  \texttt{ren@ece.ucr.edu} \\
}

\begin{document}
\maketitle

\begin{abstract}

Traffic optimization challenges, such as load balancing, flow scheduling, and improving packet delivery time, are difficult online decision-making problems in wide area networks (WAN). Complex heuristics are needed for instance to find optimal paths that improve packet delivery time and minimize interruptions which may be caused by link failures or congestion. The recent success of reinforcement learning (RL) algorithms can provide useful solutions to build better robust systems that learn from experience in model-free settings.

In this work, we consider a path optimization problem, specifically for packet routing, in large complex networks. We develop and evaluate a model-free approach, applying multi-agent meta reinforcement learning (MAMRL) that can determine the next-hop of each packet to get it delivered to its destination with minimum time overall. Specifically, we propose to leverage and compare deep policy optimization RL algorithms for enabling distributed model-free control in communication networks and present a novel meta-learning-based framework, MAMRL, for enabling quick adaptation to topology changes. To evaluate the proposed framework, we simulate with various WAN topologies. Our extensive packet-level simulation results show that compared to classical shortest path and traditional reinforcement learning approaches, MAMRL significantly reduces the average packet delivery time even when network demand increases; and compared to a non-meta deep policy optimization algorithm, our results show the reduction of packet loss in much fewer episodes when link failures occur while offering comparable average packet delivery time.
\end{abstract}



\maketitle




\section{Introduction }\label{section:introduction}
Network providers leverage traffic engineering techniques to optimize performance over operational IP networks \cite{awduche}. With the exponential growth in demand, researchers have experimented with new optimization algorithms that aim to balance utilization and availability of network resources \cite{teavar} or optimize on multiple criteria such as the number of hops, availability, or path attributes \cite{sobrinho2020routing}. As wide-area network backbones (WANs) become costly to maintain and upgrade, software-defined networking (SDN) is being used as a promising method to maximize routing performance but needs to calculate and optimize routes globally to users \cite{b4,swan}. 

Optimizing for network performance includes measuring bandwidth, jitter, or latency over resource links, where a poorly designed infrastructure can lead to slow performance and potentially increased packet loss. Apart from meticulously designed heuristics needed to develop an optimization algorithm, researchers would analyze offline models of the network topology and traffic demand matrix to infer the best paths between source-destination pairs. This approach, as a planning tool, leads to many limitations such as (1) difficult to optimize in few minutes as networks grow from 10s to 100s of routers, and (2) the dynamic traffic demand matrix would require recalculation every few days as some links become congested and possibly fail.

Recent breakthroughs in deep learning techniques leveraging data-driven learning have identified successful and simple solutions to complex online decision-making problems such as playing games \cite{alphago}, resource management \cite{mao2016resource} and learning inherent traffic patterns \cite{greguric2020application}. Particularly deep reinforcement learning (RL) uses agents to learn, through interactions with the environment by trial and error, their optimal actions via experiences and feedback. Here agents can slowly modify behavior through interactions without knowing the accurate mathematical model of the environment. In path optimization problems, RL has a natural application by exploring different routing policies, gathering statistics about which policies maximize performance functions, and learning over time the best policy on which route to take. Examples of similar techniques in path optimization include two main approaches \cite{valadarsky2017learning} such as optimizing routing configurations by predicting future traffic conditions depending on past traffic patterns or optimizing routing configurations based on the number of feasible traffic scenarios to improve performance parameters. 
Modern communication networks have become very challenging mainly due to the following two reasons \cite{boutaba2018comprehensive}. 
Firstly, communication networks have become very complicated and highly dynamic, which makes them hard to model and control. For example, in vehicular and ad hoc networks, nodes frequently move, and link failures might occur during working hours, which might result in topology changes \cite{govindan2016evolve,hong2018b4}. Second, as the scale of networks continue to multiply, a central controller may be costly to install and slow to configure and be robust to malicious attacks \cite{simplicio2010survey,al2015application}. Therefore, there is a need to develop innovative ways in which traffic routing does not rely on accurate mathematical models and can be managed in a distributed manner. Examples of distributed path planning such as ant colony optimization and swarm approaches have shown success in static environments, but still need more learning for dynamic environments \cite{Schanemann2007}.

In this paper, we investigate the above issues and evaluate if deep reinforcement learning can provide an optimal, adaptable and distributed solution to path selections as network load increases, and especially if the topology changes such as link failing or becoming congested. In order to design an optimal path selection for various network topologies and network loads, we design a deep policy-based meta-learning algorithm (MAMRL) and evaluate its performance in various simulated WAN topologies. MAMRL can optimize multiple objectives packet loss and packet delivery time. Our preliminary results show that MAMRL can perform better than the shortest possible route algorithms, especially as network loads increase and congestion is possible. Modelling the network as a multi-agent complex system, representing each router as an agent, we can demonstrate that MAMRL can learn and perform with each router in a distributed manner, allowing future work for online traffic engineering on devices.


\subsection{Motivation and Contribution}

In this work, we investigate the challenge of how can one build an adaptive network routing controller that continue to provide optimum network performance, even when the topology changes. For this we model the challenge as a path optimization technique that can be adaptive to various network load and topology changes, via novel deep reinforcement learning. Modern communication networks are highly dynamic and hard to model and predict, therefore, we aim to develop a novel \emph{experience-driven} algorithm that can learn to select paths from its experience rather than an accurate mathematical model. Additionally, due to difficulties in gathering information from widely distributed routers, we design a distributed optimization framework to learn the local optimal strategy. 

Our specific contributions are:
\begin{itemize}
    \item \textbf{Policy-based RL learning performs well at high network load.} Using a policy-based deep reinforcement learning method, we train the model at a variety of network loads and save the optimal neural network. Once deployed, our trained neural network can perform superiorly at high network load compared to value-based RL learning. 
    \item \textbf{Optimizing for multiple criteria.} Our neural networks are optimized for multi-objective optimization for both packet delivery time and packet loss on the network links. We design an appropriate reward (utility) function, which well represents the preference of the network controller, to minimize both packet delivery time and packet loss when link failures occur. 
    \item \textbf{Quick adaptation to link failures.} Our proposed MAMRL framework aims to make good online decisions under the guidance of powerful Deep Neural Networks (DNNs). In addition, by leveraging the model-agnostic meta-learning technique \cite{MAML}, our neural networks can quickly alternate paths to minimize both packet delivery time and packet loss when link failures occur. 
    \item \textbf{Calculate optimal packet routes based on limited local observation
    for future on-device Traffic Engineering research.} Our neural network model is deployed per multiple agents to represent multiple routers, allowing each router agent to learn and optimize the traffic routing based on their local information. To achieve this, we leverage a dynamic consensus estimator \cite{consensus} to diffuse local information and estimate global rewards, still achieving the best average packet delivery time. 

\end{itemize}

We develop a fully distributed multi-agent meta-reinforcement learning (MAMRL) for a  packet routing problem, where each router agent aims to find the correct adjacent routers to send their packets to minimize the overall average packet delivery time and avoid packet loss. We demonstrate our results via extensive packet-level simulations representative of WAN network topologies (ATT, Geant, B4), showing that MAMRL significantly outperforms several baseline algorithms in high network load.




\section{Background}

\subsection{Deep Reinforcement learning}\label{section:RL}
Reinforcement learning is concerned with how an intelligent agent learns a good strategy from experimental trials and relative feedback received. With the optimal strategy, the agent is capable to actively adapt to the environment to maximize cumulative rewards. Almost all the deep RL problems can be framed as Markov Decision Processes (MDPs), which consists of four key elements $\langle \mathcal{S}, \mathcal{A}, P$, $R\rangle$. More specifically, at each decision epoch $t$, the intelligent agent can stay in state $s_t$ that belongs to the state space $\mathcal{S}$ of the environment, and choose to take an action $a_t$ that belongs to the action space $\mathcal{A}$ to switches from one state to another. The probability that the process moves into its new state $s_{t+1}$ is given by the state transition function $P(s_{t+1}|s_t,a_t)$. Once an action is taken, the environment delivers a reward $r$ as feedback. Figure \ref{overallrl} shows the general process of reinforcement learning (the definition of policy will be given below).

  \begin{figure}[h]
  \centering
  \includegraphics[width=0.6\linewidth]{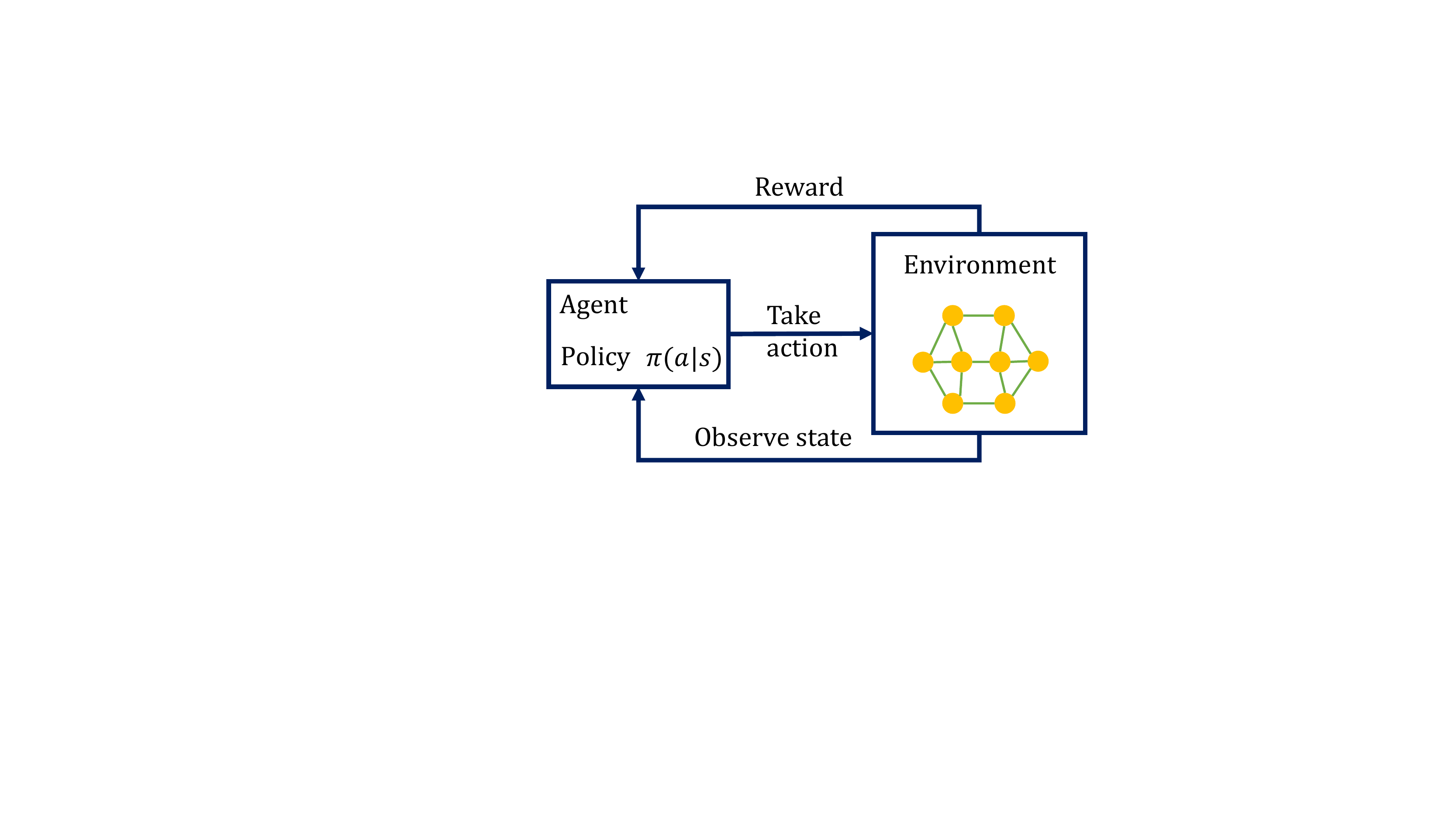}
  \caption{\label{overallrl} A global reinforcement learning agent learning network states.}
\end{figure}

There are two key functions in RL: \textbf{policy function} $\pi(a_t|s_t)$ and \textbf{value function} $V_{\pi}(s_t)/Q_{\pi}(s_t,a_t)$. The policy is a mapping from state $s_t$ to action $a_t$ and tells the agent which action $a_t$ to take in state $s_t$. For example, in the path optimizing problem, the policy is the router's strategy that finds the best adjacent router to send out current packets given the current utilization state of the communication network. The state value function $V_{\pi}(s_t)$ measures how rewarding a state is under policy $\pi$ by a prediction of future reward. Similarly, the action-state value function $Q_{\pi}(s_t,a_t)$ tells, for a given policy, what the expected cumulative reward of taking action $a_t$ in state $s_t$ is. 

The goal of RL is to find the optimal policy that achieves optimal value functions: $\pi^{*} = \text{argmax}_{\pi} V_{\pi}(s_t) = \text{argmax}_{\pi} Q_{\pi}(s_t,a_t)$. Traffic engineering is a natural application of RL by exploring with different routing policies, gathering statistics about which policies maximize the utility function, and learning the best policy accordingly. 

\textbf{Value-based algorithms versus Policy-based algorithms:} Value-based RL algorithms attempt to learn the tabular or approximation of the state-action value $Q_{\pi}(s,a)$ and selects the action based on the maximal value function of all available actions for a given state. For example, the Q-routing algorithm enables the routers to restore Q-values as the estimate of the negative transmission time between that router and the others. To shorten the average packet delivery time, routers will choose the action with maximal Q-values. Policy-based RL algorithms instead learn the policy directly with a parameterized function concerning $\theta$, $\pi_{\theta}(a_t|s_t)$ and train the policy to maximize the expected cumulative reward function. Policy-based algorithms can learn stochastic policies. It is worthwhile to note that stochastic means stochastic in some action-state pairs where it makes sense. Usually, value-based algorithms, which choose the actions with the maximal values, can only follow deterministic policies or stochastic policies with predetermined distributions. That is not quite the same as learning the real optimal stochastic policy. Since the current communication networks are highly dynamic and stochastic, we can expect that the policy-based RL algorithms perform superiorly to the value-based RL algorithms for certain scenarios, where the optimal policy is stochastic. In Section \ref{section:PG}, we will introduce more details about the policy-based RL algorithms. 

In this work, to enable high-dimensional state representations
(such as action histories), we consider deep RL algorithms, which adopt deep neural networks to approximate the policy functions. Here the policy parameters $\theta$ are the weights of the deep neural networks. 





\subsection{Performance under Partial Observability}\label{section:PG}
Figure \ref{overallrl} shows the general process of reinforcement learning, where the agent is able to observe the global information of the environment. As stated in Section \ref{section:introduction}, in this work, we consider a path optimization problem in the distributed network environment, indicating that each router only has access to its own information and the information received from its adjacent routers. It follows that the path optimization problem can be modeled as a multi-agent partially observable Markov decision process (POMDP). A POMDP, referred as $\mathcal{M}$, for $n$ routers is defined by a tuple $\langle \mathcal{S}, \{\mathcal{O}^i\}_{i\in\mathcal{V}}, \{\mathcal{A}^i\}_{i\in\mathcal{V}}, P$, $\{R^i\}_{i\in\mathcal{V}}\rangle$, where $S$ and $P$ carry the same meaning as those in Section \ref{section:RL} and $\mathcal{V}$ denotes the set of all routers. $\mathcal{O}^i$, $\mathcal{A}^i$, and $R^i$ are the local observation space, local action space and local reward function of router $i$, respectively. Then we have $\mathcal{A} = \Pi_{i=1}^n \mathcal{A}^i$ is the joint action space of all routers. Each router only has access to a private local observation correlated with the state $o^i_t$. To choose actions, each router $i$ uses a stochastic parametric policy $\pi_{\theta^i}^i: \mathcal{O}^i \times \mathcal{A}^i\to [0, 1]$, where $\pi^i_{\theta^i}(a^i_t|o^i_t)$ represents the probability of choosing action $a^i_t$ at observation $o^i_t$. Thus, the joint policy of all routers $\pi_{\theta}:\mathcal{S}\times \mathcal{A}\to[0,1]$ satisfies $\pi_{\theta}(a_t|s_t)=\Pi_{i=1}^n \pi^i_{\theta^i}(a^i_t|o^i_t)$. For a given time horizon $H$ we define the trajectory $\tau :=(s_0,a_0,\cdots,s_H,a_H,s_{H+1})$ as the collection of state action pairs ended at time $t=H$. The probability distribution of the initial state is denoted by $\rho(s_0)$. In the path optimization problem (cooperative multi-agent problem), the collective objective of all the routers is to collaboratively find policies $\pi_{\theta^i}^i$ for all $i\in\mathcal{V}$ that maximize the globally expected trajectory reward over the whole network. The goal of all routers is as follows,
\begin{equation}\label{loss}
    \max\limits_{\theta^i, i\in\mathcal{V}} J(\theta) = \mathbb{E}_{\tau}\left[R(\tau)\right],
\end{equation}
where 
$$
R(\tau) = \sum\limits_{t=0}^{H} r^i_t=\sum\limits_{t=0}^{H}\left(\tilde{r}^i_t+ \frac{1}{n}\sum\limits_{i=1}^n\hat{r}^i_t\right),
$$
and $r^i_t$ denotes the reward needed by router $i$ at time $t$. $r^i_t$ consists of two parts: 1) $\tilde{r}^i_t$ denotes the reward signal based solely on individual behavior, and 2) $\frac{1}{n}\sum_{i=1}^n\hat{r}^i_t$ denotes the reward signal based on global behavior. Note that only $\tilde{r}^i_t$ and $\hat{r}^i_t$ can be known by router $i$ in a partially observable environment.

As stated in Section \ref{section:RL}, in this work, we investigate how deep policy optimization algorithms work in path optimization problems. The main idea is to directly adjust the parameters $\theta_i, i\in\mathcal{V}$ of the policies in order to maximize the objective in \eqref{loss} by taking steps in the direction of $\nabla_{\theta_i}J(\theta_i,\cdots,\theta_n)$. For POMDP, the gradient of the expected return for router $i$ can be written\footnote{The derivation of Equation \eqref{gradient} is provided in Appendix \ref{appendix:gradient}.} as,
\begin{equation}\label{gradient}
   \nabla_{\theta^i} J(\theta^i) = \mathbb{E}_{\tau}\left[\sum\limits_{t=0}^H\nabla_{\theta^i} \log \pi^i_{\theta^i}(a^i_t|o^i_t) R(\tau)\right].
\end{equation}
Note that with only local information, function $R(\tau)$ cannot be well estimated since the estimation requires the reward $\hat{r}^i$ of all routers. In this work, we propose to use a dynamic consensus algorithm to estimate $R(\tau)$ using only local information, described in Section \ref{section:design}.  
\subsection{Model-agnostic Meta-learning}\label{section:MAML}

\begin{figure}[h]
  \centering
  \includegraphics[width=0.75\linewidth]{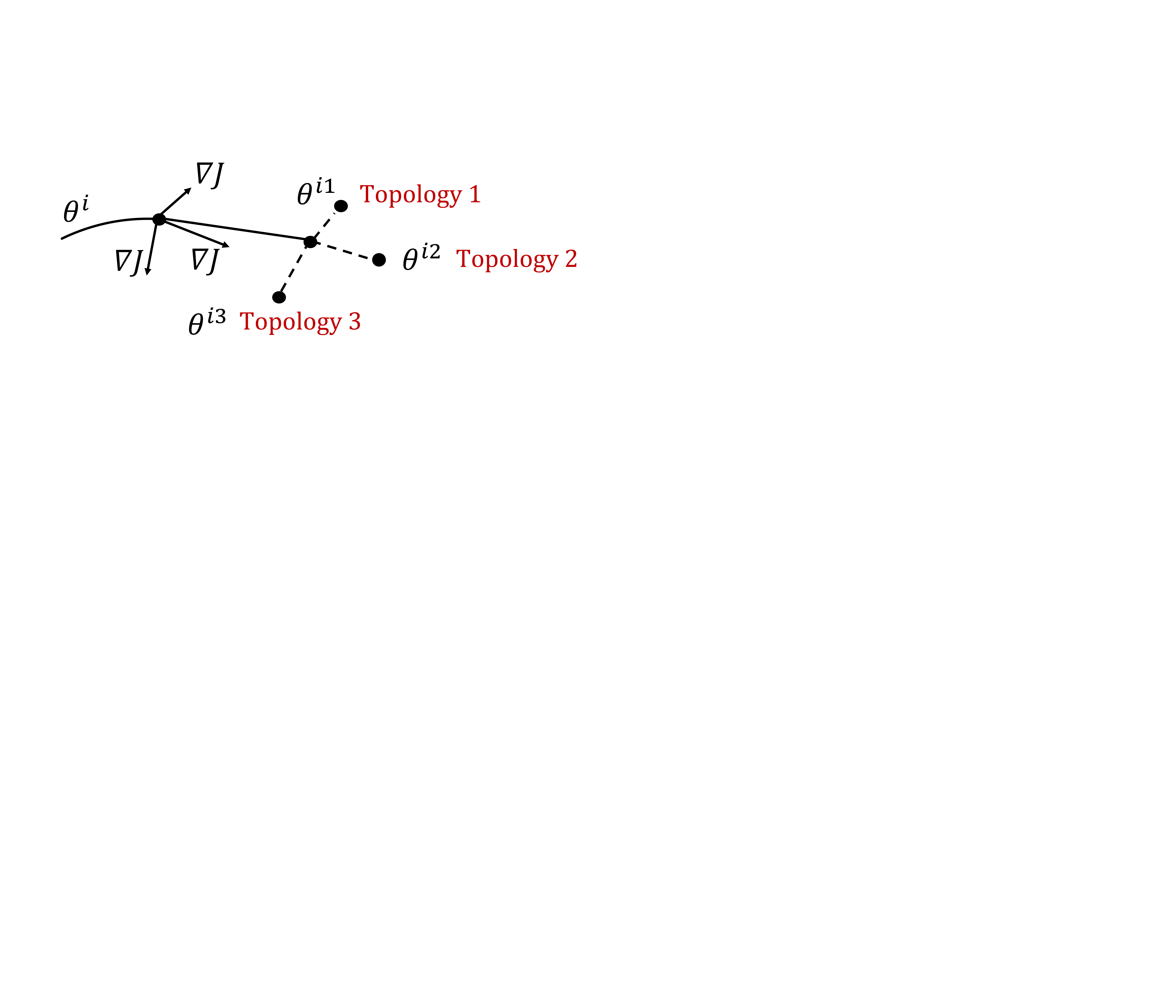}
  \caption{\label{maml}The framework of model-agnostic meta learning.}
\end{figure}
In this work, we consider path optimization in the presence of link failures. It follows that, once there is a link failure, the state transition function of the environment changes accordingly, indicating that a new POMDP occurs. 
Let $\mathcal{M}_0$ denote the Markov process modeled by the full network environment (no link failures) and $\mathcal{M}_k$, where $k>0$, denote the Markov process modeled by the network environment with different link failure scenarios. Suppose that the distribution of all POMDPs follows $\eta(\mathcal{M})$. To make the routing algorithm adapt to link failures (different POMDPs) quickly, we leverage the model-agnostic meta-learning  \cite{MAML} into the policy optimization algorithms. Meta-reinforcement aims to learn an algorithm that can quickly learn optimal policies in $\mathcal{M}_k$ drawn from a distribution $\eta(\mathcal{M})$ over a set of Markov decision processes. Our approach trains a well-generalized parametric policy initialization that is close to all the possible environments (POMDPs), such that it can quickly improve its performance on a new environment with one or a few vanilla policy gradient steps (see Figure \ref{maml}). The meta-learning objective can be written as:
\begin{equation}
\begin{aligned}
    &\max\limits_{\theta} \quad E_{\mathcal{M}\sim \eta(\mathcal{M}),\tau\sim p(\tau|\pi_{\hat{\theta}})} \left[R(\tau)\right]\\
    &\text{s.t.}\quad \hat{\theta} =\theta + \alpha \nabla_{\theta}\mathbb{E}_{\tau\sim p(\tau|\pi_{\theta})}[R(\tau)],
\end{aligned}\end{equation}
where $\alpha >0$ is the learning rate and $p(\tau|\pi_{\theta})$ represents the distribution of trajectory $\tau$ given policy $\pi_{\theta}$. Model-agnostic meta-learning attempts to learn an initialization $\theta^*$ such that for any environment $\mathcal{M}_k$ the policy attains maximum performance after a few policy gradient steps. 

\section{Design: MAMRL Approach}\label{section:design}

\subsection{Model}
In the packet routing problem, packets are transmitted from a source to its destination through intermediate routers and available links. The mathematical model is given below.

\textbf{Environment}. We consider a possibly time-varying communication network environment, which is characterized by an undirected graph $\mathcal{G}_t = (\mathcal{V},\mathcal{E}_t)$, where $\mathcal{V} = \{1,\cdots, n\}$ is a set of routers and $\mathcal{E}_t \subseteq \mathcal{V}\times \mathcal{V}$ are transmission links between the routers at time $t$. The bandwidth of each link is limited and packet loss might occur when the size of the packet to be transmitted is greater than the link's capacity. The communication network is possibly time varying since link failures might happen during working hours. When the link failure happens, the capacity of the link becomes zero. Each router $i$ has a set of neighbor routers denoted by $\mathcal{N}_i(t) = \{j \in \mathcal{V}: (i, j) \in \mathcal{E}_t\}$.

\textbf{Routing}. Packets are introduced into the network with a node of origin and another node of destination. They travel to their destination nodes by hopping on intermediate nodes. Each router only has one local port/queue used to store traffic. The queue of routers follows the first-in-first-out (FIFO) criterion. The node can forward the top packet in its local queue to one of its neighbors. Once a packet reaches its destination, it is removed from the network.

\textbf{Objective}. The packet routing problem aims at finding the optimal transmission path between source and destination routers to minimize the average packet delivery time, which is the sum of queuing time and transmission time while preventing packet loss when link failures happen.



\subsection{RL Formulation}


Our standard RL setup consists of multiple router agents interacting with an environment (communication networks) in discrete decision epochs. We investigate the deep policy optimization algorithm to address packet routing in a partially observable network environment. To make the router controllers adapt to link failures more quickly, we leverage the model-agnostic meta-learning technique to learn the well-generalized policy initialization. 


\begin{figure*}[h]
  \centering
  \includegraphics[width=0.9\linewidth]{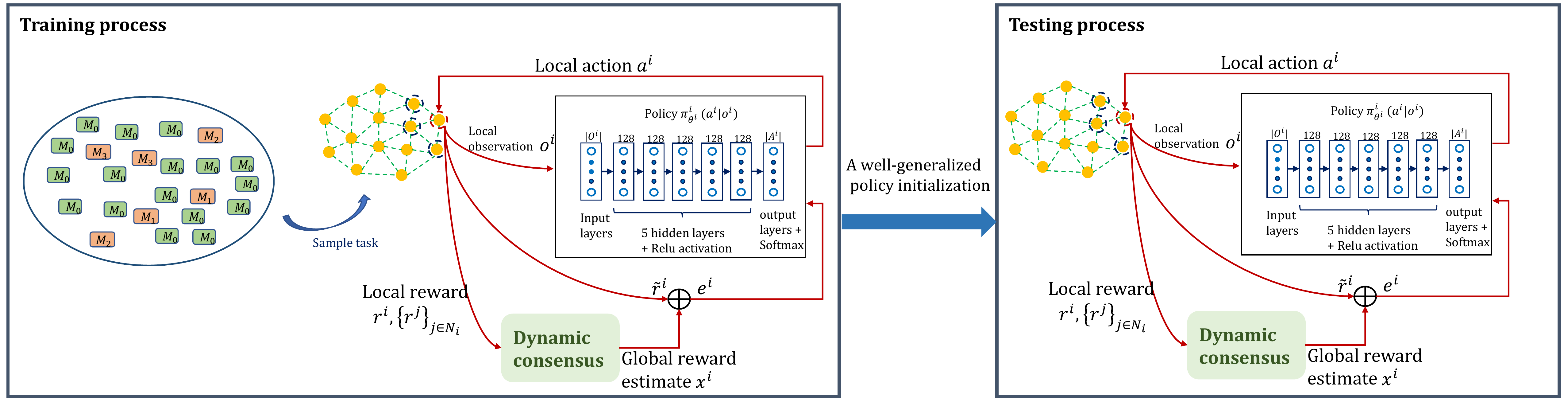}
  \caption{\label{diagram}MAMRL framework.}
\end{figure*}
Figure \ref{diagram} shows the MAMRL setup (training and testing process) per router. In the testing process, each router uses the deep policy optimization algorithm coupled with the dynamic consensus estimator to learn the optimal policy. In order to let the routers adapt to topology changes quickly, the policy of each router is initialized using the well-generalized policy initialization, which is the output of the training process. The training process follows the traditional model-agnostic meta-reinforcement learning framework. The basic idea is letting the network controller encounter multiple link failures in the training process. It can use this experience to learn how to adapt if similar situations occur while deployed. In Figure \ref{diagram}, $\mathcal{M}_0$ denotes the Markov process modeled by the full network environment (no link failures) and $\mathcal{M}_k$, where $k>0$, denotes the Markov process modeled by the network environment with link failures. In the training process, the network controller collects data samples from all possible network environments according to the distribution $\eta(\mathcal{M})$. However, the traditional design of model-agnostic meta-learning mainly focuses on single-agent (centralized) reinforcement learning problems. How to solve a multi-agent RL problem using model-agnostic meta-learning in a distributed manner is rarely studied. In this work, we aim to train and execute the network controller in a distributed manner. As shown in Figure \ref{diagram}, each router has an independent policy model that is represented by a deep neural network. The core of the proposed control framework is letting each router run a deep reinforcement learning algorithm to find the best action at each decision time instant, using only local information and local interaction. Since the routers aim to minimize the average packet delivery time of the whole network, each router needs to feed the global packet delivery time into its policy model as feedback/reward. To achieve this goal, we leverage the dynamic consensus algorithm to estimate the global reward function, which is described below, 


The policy optimization algorithms aim to find the best policy parameters that produce the highest long-term expected return using gradient ascent. The gradient of the long-term expected return for the parameters of each router's policy is defined in Equation \eqref{gradient}. However, with only local information, function $R(\tau)$ cannot be well estimated since the estimation requires the rewards $\hat{r}^i$ of all routers $\forall i \in \mathcal{V}$. This motivates our consensus-based policy gradient algorithm that leverages the communication network to diffuse the local information, fostering collaboration among routers. We adapt the following dynamic consensus algorithm \cite{consensus} into the policy optimization method. 

\begin{equation}\label{dynamic_consensus}
\begin{aligned}
x^i_{t} &= \hat{r}^i_t - y^i_t,\\
y^i_{t+1}& = \beta\sum\limits_{j\in\mathcal{N}_i}(x^i_{t}-x^j_{t}) + y_{t}^i,
\end{aligned}
\end{equation}
where $0<\beta<1$ is the control gain, $x_t^i$ and $y_t^i$ are local estimators, and $\mathcal{N}_i$ denotes the neighbor sets of router $i$. It can be proved that $x_t^i$ converges to the vicinity of $\frac{1}{n}\sum_{i=1}^n\hat{r}^i_t$ within a few time steps. It is worthy to mention that only local information is used in the designed estimator Equation \eqref{dynamic_consensus}. 

We develop the following policy optimization method for POMDP,
\begin{equation}\label{algorithm}
   \nabla_{\theta^i} \bar{J}(\theta^i) = \mathbb{E}_{\tau}\left[\sum\limits_{t=0}^H\nabla_{\theta^i} \log \pi^i_{\theta^i}(a_t^i|o_t^i) \bar{R}^i(\tau)\right],
\end{equation}
where
\begin{equation}\label{tildeQ}
\bar{R}^i(\tau) =\sum\limits_{l=0}^H e^i_{l}=\sum\limits_{l=0}^H \tilde{r}_l^i+x^i_{l} \approx \sum\limits_{l=0}^H\left(\tilde{r}_l^i+ \frac{1}{n}\sum\limits_{i=1}^n\hat{r}^i_{l}\right).
\end{equation}
Here, $e^i_l$ denotes the sum of the local reward signal $\tilde{r}^i_l$ and global reward estimate $x^i_{l}$. And $x^i_l$ is obtained by the dynamic consensus estimator designed in Equation \eqref{dynamic_consensus}. Note that both $\tilde{r}^i$ and $x^i$ can be obtained locally.

We build the deep neural network with one input layer, five hidden layers of size 128 with ReLU, and one output layer with Softmax (see Figure \ref{diagram}). As shown in Figure \ref{diagram}, at each decision epoch $t$, each router $i$ provides the local observation $o^i_t$ to the policy model $\pi_{\theta^i}$ and gets the action $a^i$ back. Router $i$ performs action $a^i_t$ and switch to a new state. Then router $i$ feeds the local reward $\tilde{r}^i_{t+1}$ and global reward estimate $x^i_{t+1}$, which is the output of the dynamic consensus estimator, to the policy model and the policy model $\pi_{\theta^i}$ updates its weights $\theta^i$ with respect to the received reward estimate $e^i$. It is worthwhile to mention that to update the policy in the direction of greater cumulative reward using Equation \eqref{algorithm}, only local information $o^i_t$, $a^i_t$ and $e^i_t$ are required. By integrating model-agnostic meta-learning and the proposed multi-agent policy optimization algorithm, MAMRL for packet routing problem, where both training and execution process is distributed. These are shown in Algorithms \ref{MAMMAL_train} and \ref{MAMMAL_test}. 

\begin{algorithm}
\SetAlgoLined
\textbf{Input}: $\eta(\mathcal{M})$: distributions of network environments\;
\textbf{Input}: $\alpha$: step size hyper-parameter\;
randomly initialize $\theta^i, i\in \mathcal{V}$\;
 \While{not done}{
  sample batch of environments $\mathcal{M}_k \sim \eta(\mathcal{M})$\;
  \For{all $\mathcal{M}_k$}{
    \For{all routers $i\in\mathcal{V}$}{Sample $K$ trajectories $D^i=\{(o_0^i,a_0^i,e_0^i,\cdots,o^i_H,a^i_H,e^i_H )\}$ using $\pi_{\theta^i}^i$ and Equation \eqref{dynamic_consensus} in $\mathcal{M}_k$\;}
    \For{all routers $i\in\mathcal{V}$}{Evaluate $\nabla _{\theta^i} \bar{J}(\theta)$ using $D^i$ based on Equation \eqref{algorithm}\;
   Compute adapted parameters with gradient descent: $\hat{\theta}^i=\theta^i -\alpha \nabla_{\theta^i} \bar{J}(\theta)$\;}
   
   \For{all routes $i\in\mathcal{V}$}{
   Sample $K$ trajectories $\hat{D}^i=\{(o_0^i,a_0^i,e_0^i,\cdots,o^i_H,a^i_H,e^i_H )\}$ using $\pi_{\hat{\theta}^i}^i$ and Equation \eqref{dynamic_consensus} in $\mathcal{M}_k$\;}
   }{
   \For{all routes $i\in\mathcal{V}$}{
   Evaluate $\nabla _{\hat{\theta}^i} \bar{J}(\hat{\theta})$ using $\{\hat{D}^i\}_{i\in\mathcal{V}}$ based on Equation \eqref{algorithm}\;
   Update $\theta^i$ with gradient descent:
   $\theta^i=\theta^i -\alpha \nabla_{\hat{\theta}^i} \bar{J}(\hat{\theta})$\;
   using $\hat{D}^i$ based on Equation \eqref{algorithm}\;}
  }
 }
  \textbf{Return} $\theta^i,\ i\in\mathcal{V}$ as parameter initialization.
 \caption{\label{MAMMAL_train} Multi-agent meta reinforcement learning algorithm  (MAMRL train time)}
\end{algorithm}

\begin{algorithm}
\SetAlgoLined
\textbf{Input}: A dynamic network environment with possible task distribution $\eta(\mathcal{M})$\;
\textbf{Input}: $\alpha$: step size hyper-parameter\;
\textbf{Input}: Learned parameter initialization $\theta^i, i\in \mathcal{V}$\;
 \While{not done}{
 \If{Link failure is True}{$\theta_t^i \leftarrow \theta^i$, $i\in\mathcal{V}$}
 \For{all routers $i\in\mathcal{V}$}{Sample $K$ trajectories $D=\{(o_0^i,a_0^i,e_0^i,\cdots,o^i_H,a^i_H,e^i_H )\}$ using $\pi_{\theta^i}^i$ and Equation \eqref{dynamic_consensus}\;}
 Evaluate $\nabla _{\theta^i} \bar{J}(\theta^i)$ using $D$ based on Equation \eqref{algorithm}\;
 \For{all routers $i\in\mathcal{V}$}{
 Compute adapted parameters with gradient descent: $\theta_{t+1}^i=\theta_{t}^i -\alpha \nabla_{\theta^i} \bar{J}(\theta^i)$\;}
 }
 \caption{\label{MAMMAL_test} Multi-agent meta reinforcement learning algorithm (MAMRL test time)}
\end{algorithm}
\subsection{Design of State, Action space and Rewards}


We design the local observation $o^i_t$, local action $a^i_t$ and local estimation of the reward function $e^i_t$ below,
\begin{itemize}
    \item \textbf{Observation of router $i$, $o_i$}: 1) destination router of first packet in the local queue; 2) the last ten step actions taken by router $i$; 3) the address of the router which has the longest queue among all the neighbor router of router $i$.

    \item \textbf{Action of router $i$, $a_i$}: next hop of current packet in the queue.

    \item \textbf{Reward estimate of router $i$, $e^i$}: sum of $\tilde{r}^i$ and $x^i$, where $\tilde{r}^i$ is negative number of packet loss occurred at router $i$ and $x^i$ is the estimate of $\frac{1}{n}\sum_{j=1}^n\hat{r}^j$ using Equation \eqref{dynamic_consensus}. Here, $\hat{r}^j$ denotes the negative average delivery time of all the packets delivered to router $j$.
\end{itemize}
Note that the design of state space and reward
is critical to the success of a deep reinforcement learning method. Our design of the state space captures key components of the network environment. For our design of the reward function, element $\frac{1}{n}\sum_{j=1}^n \hat{r}^j$ is introduced to minimize the average packet delivery time of the whole network, element $\tilde{r}^i$ is included to minimize the packet loss occurred at router $i$ in the presence of link failures. Note that in our design $e^i = x^i + \tilde{r}^i$, where $\tilde{r}^i$ is the negative number of packet loss occurred only at router $i$ but $x^i$ is the estimate of the negative average delivery time of the whole network environment. The reasons are summarized below.

\textbf{Optimizing for packet delivery time:} To achieve this goal, all the routers need to collaboratively find the best paths to reroute the traffic. And the delivery time of the packets that are delivered to router $i$ is determined by the decisions of all the intermediate routers. That is, packet delivery time is a signal based on global behavior, it is not enough for router $i$ to only know the delivery time of the packets delivered to itself.

\textbf{Optimizing for link failures:} Although this goal also involves reading packet loss in the whole network, the link failures have little effect on the routers that are not directly connected to the failed links. Therefore, in our design, we only provide the packet loss that occurred at router $i$ to the policy $\pi_{\theta^i}$ as the feedback.




\section{Evaluation}
We conduct extensive simulations to evaluate the performance of the proposed MAMRL framework in a path optimization problem with static topologies and topologies with possibly failed links. 
\begin{table}
    \centering
    \begin{tabular}{c|c|c}
         \hline
         Topology Name & Number of nodes & Number of edges  \\
         \hline 
         B4 & 12 & 19\\
         Geant & 21 & 32\\
         ATT & 25 & 56\\
         \hline
    \end{tabular}
    \caption{Network topologies used in our evaluations.}
    \label{table:topo}
\end{table}
We evaluate the results to, \begin{itemize}
    \item Benchmark the RL techniques against standard path optimization approaches and other RL approaches,\item Examine how robust the MAMRL approach is under link failures, and \item Show how quickly MAMRL adapts and reroutes packets to alternate paths achieving better performance.\end{itemize} 

\subsection{Experiment Settings}
The simulation runs are performed on three network topologies, B4, Geant, and ATT network. See Table \ref{table:topo} for a specification of network sizes.
The B4 and ATT topologies (link capacities) and their traffic matrices (packet size) were obtained from the authors of Teavar \cite{teavar}. The Geant topology is the European Research network providing connectivity to science experiments across Europe and US labs (www.geant.org).  

To model the packet arrival, a discrete event network simulator is developed, based on Open AI gym \footnote{https://github.com/esnet/daphne-public/tree/master/MAMRL-TE}. Packets are introduced into the network with a node of origin and another node of destination. The packet arrives according to the Poisson process of rate $\lambda$. They travel to their destination node by hopping on intermediate nodes. Each router only has a one local port/queue used to store traffic. The queue of routers follows the FIFO criterion. In each time unit, the node forwards the top packet in its local queue to one of its neighbors. Once a packet reaches its destination, it is removed from the network environment. The bandwidth of each link is limited and packet loss might occur when the size of the packet to be transmitted is greater than the link's capacity. When the link failure happens, the capacity of the link becomes zero. 

In the experiments, we choose the step size $\alpha$ as $0.01$. In addition, we use trust-region policy optimization (TRPO) \cite{trpo} as the meta-optimizer and the standard linear feature baseline \cite{baseline} is used. 

\subsection{Impact of Increasing Network Load}
We first test the MAMRL algorithm with static topologies (no link failures). We compare with the classical shortest path algorithm and two existing RL-based routing algorithms: 
\begin{itemize}
    \item Shorted path algorithm (SPA) \cite{spa}: a traditional packet routing algorithm.
    \item Q-routing \cite{boyan1994packet}: a value-based multi-agent reinforcement learning algorithm.
    \item Policy gradient (PG) \cite{peshkin2002reinforcement}: a policy-based multi-agent reinforcement learning algorithm.
\end{itemize}

In the experiments, the episodes terminate at the horizon of $H = 500$. After 10000 training episodes, we restored the well-trained models to compare their performance in a new test environment where packets were generated at the corresponding network load level. Note that the SPA does not need training and can be applied to test directly. We tested the network on loads
ranging from 0.005 to 0.5 and measured the average packet delivery time of several episodes in the testing process to compare with the results given by the above-mentioned baseline controllers. The load corresponds to the value of $\lambda$ of the Poisson arrival process for the average number of packets injected per unit time.

\begin{figure*}[htb]
    \centering
    \begin{subfigure}[b]{0.32\textwidth}
        \centering
        \includegraphics[width=0.95\linewidth]{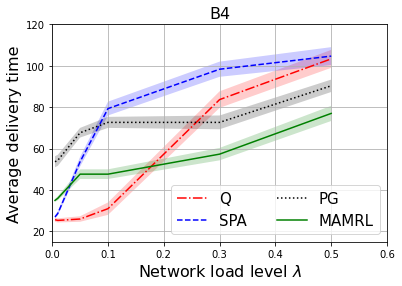}
        \caption{B4.}
    \end{subfigure}
    \begin{subfigure}[b]{0.32\textwidth}
        \centering
        \includegraphics[width=0.95\linewidth]{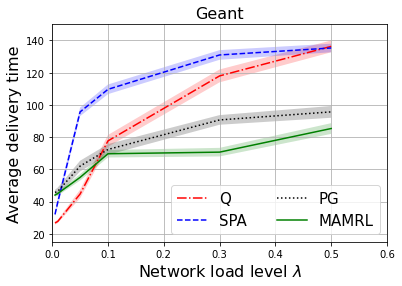}
        \caption{Geant.}
    \end{subfigure}
    \begin{subfigure}[b]{0.32\textwidth}
        \centering
        \includegraphics[width=0.95\linewidth]{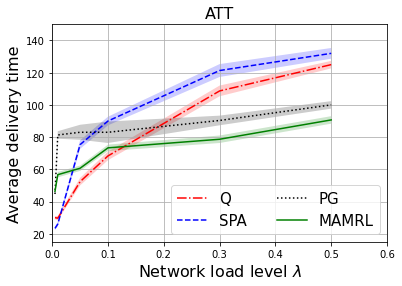}
        \caption{ATT.}
    \end{subfigure}
    \caption{Comparing average packet delivery time as load increases.}
\label{results_static}
\end{figure*}

The average packet delivery time results versus different network load are shown in Figure \ref{results_static}. Under conditions of low load for all the three topologies, MAMRL is slightly inferior to Q-routing and SPA. As the load increases, the MAMRL performs much better than the baseline algorithms. On the B4 topology, when the traffic load is high (i.e., $\lambda = 0.5$), MAMRL reduces the average packet delivery time by $25\%$, $24\%$, and $14\%$, respectively, compared to SPA, Q-routing, and policy gradient algorithms. On the Geant topology, when the traffic load $\lambda = 0.5$, MAMRL significantly reduces the average packet delivery time by $37\%$, $37\%$, and $10\%$, respectively, compared to SPA, Q-routing, and policy gradient algorithms. And on the ATT topology, when the traffic load $\lambda = 0.5$, MAMRL reduces the average packet delivery time by $33\%$, $28\%$, and $10\%$, respectively, compared to SPA, Q-routing, and policy gradient algorithms. The reason is described as follows. Under conditions of low load, there is no congestion along the route. Therefore, the deterministic policy learned by Q-routing performs as well as SPA, which is the optimal routing policy under low load. However, the routing policy learned by MAMRL is stochastic, which means that not all of the packets are sent down the optimal link. That is why the performance of MAMRL is slightly inferior to Q-routing and SPA under low load. As the load increases, the routes are getting crowded and the length of the queues is getting longer. Due to the stochastic nature of the communication network environment, the optimal policy should be stochastic under conditions of high load. This explains why MAMRL performs much better than the Q-routing and SPA controllers under high load. The results in \cite{peshkin2002reinforcement} also show that policy-based reinforcement learning algorithm performs better than value-based algorithms, especially on high flow load. However, the work in \cite{peshkin2002reinforcement} only considers a simple policy gradient algorithm for the packet routing problem. Instead, we investigate a deep policy optimization algorithm that can take much more information as its inputs, enlarging the state-action space for better policy making. The results in Figure \ref{results_static} indicate that our MAMRL algorithm achieves a shorter delivery time than a simple policy gradient algorithm.

\begin{figure*}[htb]
    
    \centering
    \begin{subfigure}[b]{0.32\textwidth}
        \centering
        \includegraphics[width=0.95\linewidth]{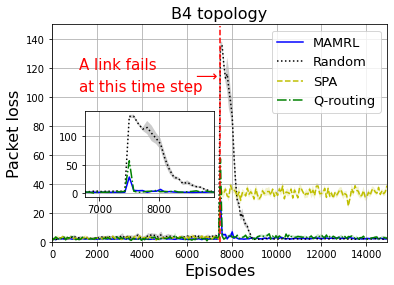}
        \caption{B4.}
    \end{subfigure}
    \begin{subfigure}[b]{0.32\textwidth}
        \centering
        \includegraphics[width=0.95\linewidth]{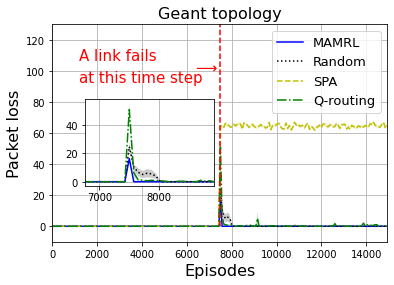}
        \caption{Geant.}
    \end{subfigure}
    \begin{subfigure}[b]{0.32\textwidth}
        \centering
        \includegraphics[width=0.95\linewidth]{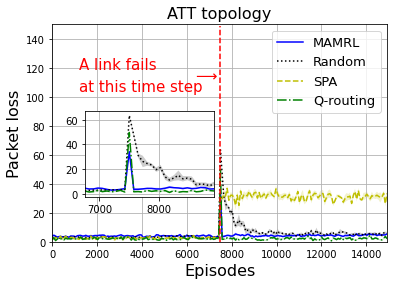}
        \caption{ATT.}
    \end{subfigure}
    \caption{Packet loss results in the presence of link failures.}
    \label{results_pl}
\end{figure*}

\subsection{Impact of Link Failures}
\begin{figure*}[htb]
    
    \centering
    \begin{subfigure}[b]{0.32\textwidth}
        \centering
        \includegraphics[width=0.95\linewidth]{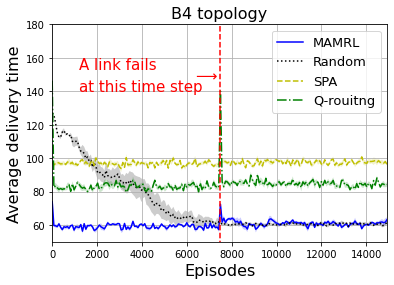}
        \caption{B4.}
    \end{subfigure}
    \begin{subfigure}[b]{0.32\textwidth}
        \centering
        \includegraphics[width=0.95\linewidth]{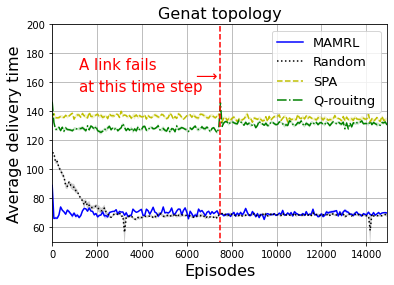}
        \caption{Geant.}
    \end{subfigure}
    \begin{subfigure}[b]{0.32\textwidth}
        \centering
        \includegraphics[width=0.95\linewidth]{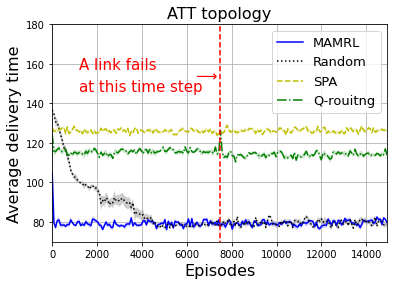}
        \caption{ATT.}
    \end{subfigure}
    \caption{Average packet delivery time results in the presence of link failures.}
    \label{results_apct}
\end{figure*}

\begin{figure*}[htb]
   
    \centering
    \begin{subfigure}[b]{0.32\textwidth}
        \centering
        \includegraphics[width=0.95\linewidth]{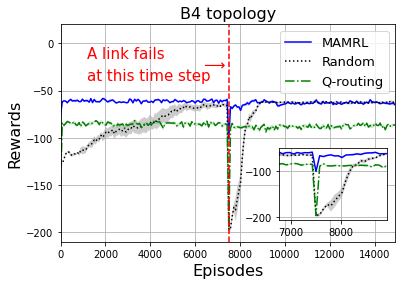}
        \caption{B4.}
    \end{subfigure}
    \begin{subfigure}[b]{0.32\textwidth}
        \centering
        \includegraphics[width=0.95\linewidth]{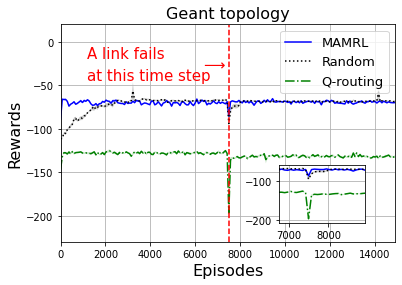}
        \caption{Geant.}
    \end{subfigure}
    \begin{subfigure}[b]{0.32\textwidth}
        \centering
        \includegraphics[width=0.95\linewidth]{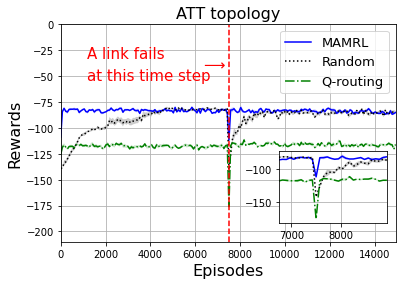}
        \caption{ATT.}
    \end{subfigure}
    \caption{Reward in the presence of link failures.}
    \label{results_reward}
\end{figure*}

We test the MAMRL algorithm in the presence of link failures with network load $\lambda = 0.3$. We let the router train and encounter all possible network environments (link failure scenarios) according to the distribution $\eta(\mathcal{M})$ and return policy parameters using Algorithm \ref{MAMMAL_train}. We restored the well-trained models in a new test environment where the links get disconnected according to the distribution $\eta(\mathcal{M})$ (We assume that only one link gets failed at one time). We compare results of using the following three controllers: (1) testing the policy from the initialization parameters obtained by MAMRL, (2) testing the policy from randomly initialized weights (called random in the following), (3) shortest path algorithm (SPA) \cite{spa}, and (4) Q-routing algorithm \cite{boyan1994packet}. Figure \ref{results_pl} show the results of the packet loss versus episodes and Figure \ref{results_apct} show the results of the average packet delivery time versus episodes. Also, we show the performance of reinforcement learning routing algorithms (MAMRL, random, Q-routing) over the three network topologies during the online learning procedure in terms of the reward. We present the corresponding simulation results in Figure \ref{results_reward}. We can make the following observations from these results. 

\begin{enumerate}
\item In Figure \ref{results_pl}, when there is a link failure, the model-based routing algorithm (i.e., SPA) witnesses a huge packet loss. The reason is that the SPA algorithm relies on previous knowledge of the network topology to make decisions. Here we assume that as the networks grows, it becomes longer to update the ISIS/OSPF protocols for link failures and update the tables. Both ISIS/OSPF use the same Dijkstra algorithm for computing the best path through the network. The other learning algorithms (MAMRL, random, Q-routing) are model-free controllers and the policy of the model-free controller. When link failure happens, the packet loss sensor will tell the RL routing controllers that there are many packet loss at the particular link. Based on our design, the packet loss hurts the reward of the RL routing controllers. To maximize the reward function, the RL routing controller will adjust their policies to improve the reward function and hence reduce the packet loss accordingly.   

\item Figure \ref{results_reward} shows how the reward value changes during online learning over the three network topologies. It is seen that when there is a link failure, for the B4 topology, Q-routing adapts to the link failure (reward values converge to the stable states) after about 30 episodes, MAMRL (our algorithm) adapts to the link failure after about 35 episodes and random algorithm (deep policy optimization with randomly initialized weights) adapts to the link failure after about 800 episodes. The results for Geant topology and ATT topology are shown in Table \ref{table:episodes}. Q-routing is based on a value-based Q-learning algorithm and is often much faster to learn a policy than policy optimization algorithms \cite{nachum2017bridging}. In this work, we propose the MAMRL algorithm which leverages model-agnostic meta-learning to help the policy optimization adapt to link failures quickly. The basic idea of MAMRL is letting the network controller encounter all possible link failures in the training process. It can then use that experience to learn how to adapt. MAMRL aims to learn a well-generalized policy initialization that is close to all possible situations of the environment. Whenever there are continual packet losses at a particular link, the MAMRL controller will reinitialize the policy models based on the pre-trained well-generalized policy initialization. It can be seen from Figure \ref{results_reward}, the MAMRL controller adapts to link failures with a speed that is comparable to the Q-routing algorithm. However, the normal policy optimization controller adapts to the link failures much more slowly. 
\end{enumerate}
\begin{table}
    \centering
    \begin{tabular}{c|c|c|c}
         \hline
         Topology Name & Q-routing & MAMRL & Random  \\
         \hline 
         B4 & 23 & 25 & 800\\
         Geant & 35 & 35 & 100\\
         ATT & 29 & 30 & 1000\\
         \hline
    \end{tabular}
    \caption{Average number of episodes used to adapt to link failures.}
    \label{table:episodes}
\end{table}

\subsection{Optimizing for Multiple Objectives}
Policy optimization algorithms use gradient descent to optimize an optimization problem. And traffic engineering aims at finding a solution to forward the data traffic to maximize a utility function. The utility function might concern a set of values. In our design, the objective is to minimize the packet delivery time and packet loss, therefore, the utility function, which corresponds to the reward function in the RL algorithms, consists of a function of packet loss and a function of packet delivery. 

In future works, we can add multiple objectives such as bandwidth utilization, latency and more, if we want the RL controller to optimize on a number of multiple parameters. 

\section{Related work}\label{literature}
Routing is the process of selecting a path for traffic in a network from the source node to a destination node. Routing algorithm selection may depend on multiple criteria, for example, performance metric (delay, link utilization), the ability to adapt to changes in topology and traffic, and scalability (should be able to support a large number of routers). In the following, we briefly review the traditional routing algorithms and reinforcement learning routing algorithms based on the above-mentioned criteria. Since in this work, we propose a kind of multi-agent reinforcement learning algorithm to solve the packet routing algorithm, a literature review on multi-agent reinforcement learning algorithms is also included in this section. 
\subsection{Traditional Packet Routing Algorithms}
There are multiple routing algorithms in the literature of traditional packet routing \cite{flooding,dijkstra,bellman}. 
Among these traditional packet routing algorithms, the shortest path algorithm is the most commonly used routing algorithm \cite{abolhasan2004review}. The shortest path algorithm aims to find the shortest path between source and destination nodes and get the packet delivered to the destination node as quickly as possible. The shortest path algorithm is regarded as the best routing algorithm on lower network load since packets can be delivered using the least amount of time along the shortest path between two nodes provided that there is no congestion along the route. However, when the network load is high, the shortest path algorithm will cause a serious backlog in busy routers. Another problem with the shortest path algorithm is that it relies on having full knowledge of the network topology to design routing algorithms and hence needs manual adjustment when topology or traffic changes happen.

\subsection{Reinforcement Learning for Routing}
Using reinforcement learning for packet routing has attracted increasing interest recently. Various reinforcement learning methods have been proposed to deal with this classical communication network problem and achieved better performances compared with traditional routing methods. 

Applications of traditional RL to solve packet routing problems started in the early 1990s with the seminal work \cite{boyan1994packet}, where Q-routing was proposed. Q-routing \cite{boyan1994packet} is an adaptive routing approach based on a reinforcement learning algorithm known as Q-learning. Q-routing routes packets based on the learned delivery times (Q values) and achieves a much smaller average delivery time compared with the benchmark shortest-path algorithm \cite{spa}. Since then, several extensions of Q-routing have been proposed, e.g., Dual Q-routing \cite{kumar1997dual}, Predictive Q-routing \cite{choi1996predictive}, Full Echo Q-routing \cite{kavalerov2017reinforcement}, Hierarchical Q-routing \cite{lopez2011simulated} and Ant-based Q-routing \cite{subramanian1997ants}. However, Q-routing is a value-based RL algorithm. That is, Q-routing is a deterministic algorithm that might cause traffic congestion at high loads and does not distribute incoming traffic across the available links. Due to the drawbacks of a value-based algorithm, some researchers begin to consider policy-based RL algorithms for packet routing problems \cite{tao2001multi,peshkin2002reinforcement}. Since policy-based RL algorithms can explore the class of stochastic policies, it is natural to expect policy-based algorithms to be superior for certain types of network topologies and loads, where the optimal policy is stochastic. In \cite{peshkin2002reinforcement}, the results show that policy-based RL algorithms perform better than value-based algorithms, especially on high flow load. These traditional reinforcement learning routing algorithms use tabular functions or simple algebraic functions to estimate the Q functions or policy functions. This is limiting for a large number of states and thus cannot take full advantage of the network traffic history and dynamics. In this work, we investigate the policy-based deep reinforcement learning algorithm and use deep neural networks to approximate the policy function. The combination of deep learning techniques with reinforcement learning methods can learn useful representations for the routing problems with high dimensional raw data input and thus achieve superior performance.  

Deep reinforcement learning is the combination of reinforcement learning and deep learning, which has been able to solve a wide range of complex decision-making tasks. However, rare works are investigating how deep reinforcement learning can be leveraged for packet routing problems since the wireless network is a multi-agent environment and the network environment is non-stationary from the perspective of any individual router. This prevents the straightforward use of experience replay, which is crucial for stabilizing deep Q learning \cite{MADDPG}. \cite{mukhutdinov2019multi} combines the Q-routing and deep Q-learning to solve the routing problem. However, the training process of the algorithm proposed in \cite{mukhutdinov2019multi} is in a centralized manner (all the routers need to share parameters), which might cause issues in real-world large-scale network environments. The authors in \cite{xu2018experience} propose to use a deep actor-critic reinforcement learning algorithm to optimize the performance of the communication network. However, the training and testing process in \cite{xu2018experience} are also in a centralized manner. Recently, distributed Deep Q-routing has been proposed in \cite{you2020toward}, where deep recurrent neural network (LSTM) has been utilized to tackle the non-stationary problem in multi-agent reinforcement learning. However, the assumptions in \cite{you2020toward} are different from those in our work. In \cite{you2020toward}, it is assumed that the bandwidth of each link equals the packet size, in which case only a single packet can be transmitted at a time. However, in our current work, we assume that the capacity of each link is fixed, in which case many packets can be transmitted at a time. We believe that our assumption is more realistic.

\subsection{Multi-agent Deep Reinforcement Learning Approaches}
There is a huge body of literature on single-agent deep reinforcement learning algorithms, where the environment stays largely stationary. Unfortunately, traditional deep reinforcement learning algorithms are poorly suited to multi-agent environments, where the environment becomes non-stationary from the perspective of any individual agent. This might cause divergence for value-based reinforcement learning and very high variance for policy-based reinforcement learning algorithms. In the literature, researchers propose multiple methods to apply reinforcement learning algorithms in multi-agent settings. To name a few, centralized training and distributed execution \cite{MADDPG,iqbal2019actor,li2019robust}, distributed training and execution under fully-observable environments \cite{zhang2018networked}, and independent Q-learning \cite{hausknecht2015deep,matignon2007hysteretic,foerster2017stabilising}. However, independent Q-learning is a value-based reinforcement learning algorithm, and in this work, we aim to investigate the policy-based reinforcement learning routing algorithm. The ideas of centralized training and fully observable state space work well when there exists a small number of agents in the communication network. With increasing the number of agents, the volume of the information might overwhelm the capacity of a single unit. To tackle this problem, one effective idea to remove the central unit and only allow the agents to share information with only a subset of agents, to reach a consensus over a variable with these agents (called neighbors) \cite{wai2018multi,zhang2016data}. In this work, we also leverage the dynamic consensus algorithm to estimate the global reward function through interactive communication among routers. Moreover, we consider packet routing in the presence of link failures, indicating that not only the local environment from the perspective of any individual router is non-stationary but also the global environment changes during the working hours. We propose to use model-agnostic meta-learning to learn a well-generalized policy initialization that is close to all possible environments such that the policy can be quickly adapted to different scenarios with a few gradient steps. This is the first time in the literature that the model-agnostic meta-learning is applied to a multi-agent reinforcement learning problem case.


\section{Discussion and Conclusions}


In this work, we propose a novel framework MAMRL that utilizes deep policy optimization and meta-learning to produce a model-free network routing controller that can perform better path optimization than standard approaches. Our experiments show that MAMRL can learn to control the communication networks from its experience rather than an accurate mathematical model. Specifically, we use deep policy optimization techniques to find optimal paths in complex WAN topologies. In order to address the difficulties in gathering information from widely distributed routers, we design a consensus-based policy optimization algorithm that can learn the local optimal strategy using only local information. Additionally, we consider path optimization problems in the presence of link failures and we leverage the model-agnostic meta-learning algorithm to make the proposed network controller adapt to link failures more quickly. We demonstrate how MAMRL improves the learning efficiency of deep reinforcement learning in multi-agent packet routing in the presence of link failures. The experiments demonstrate the effectiveness and efficiency of MAMRL for packet routing problem, compared to some baseline controllers. 

The distributed nature of MAMRL lays foundation for our future work, where we will experiment with on-device traffic engineering in physical network setups to see how well the network adapts. 


\section{Acknowledgements}
We would like to thank Dr. Manya Ghobadi for providing the data of At\&T and B4 network topologies. We would like to express our gratitude to Dr. Bashir Mohammed for his useful suggestions and critiques of this research work. This work was supported by the U.S. Department of Energy, Office of Science Early Career Research Program for `Large-scale Deep Learning for Intelligent Networks' Contract no  FP00006145.

\bibliographystyle{unsrt}  
\bibliography{main}

\begin{appendices}
\section{Appendix}\label{appendix:gradient}
Here, we provide the derivation of Equation \eqref{gradient}. All the notations carry the same meaning as those in Section \ref{section:PG}.
The probability of a trajectory $\tau=(s_0,a_0,\cdots,s_H,a_H,S_{H+1})$ given that actions come from $\pi_{\theta}$ is, 
\begin{equation}
\begin{aligned}
    P(\tau|\pi_{\theta}) &= \rho(s_0)\Pi_{t=0}^H P(s_{t+1}|s_t,a_t)\pi_{\theta}(a_t|s_t)\\
    & = \rho(s_0)\Pi_{t=0}^H P(s_{t+1}|s_t,a_t)\Pi_{i=1}^n\pi^i_{\theta^i}(a_t^i|o_t^i)
\end{aligned}\end{equation}
The log-probability of a trajectory is 
\begin{equation}
\begin{aligned}
\log P(\tau|\pi_{\theta}) &= \log \rho(s_0)+\sum_{t=0}^H\log\left[ P(s_{t+1}|s_t,a_t)\Pi_{i=1}^n\pi^i_{\theta^i}(a_t^i|o_t^i)\right]\\
&= \log \rho(s_0)+\sum_{t=0}^H\log\left[ P(s_{t+1}|s_t,a_t)\right]\\
&\qquad+\sum_{t=0}^H \sum_{i=1}^n\log\left[\pi^i_{\theta^i}(a_t^i|o_t^i)\right].
\end{aligned}
\end{equation}
The gradient of the log-probability of a trajectory is
\begin{equation}
\begin{aligned}
\nabla_{\theta}\log P(\tau|\pi_{\theta}) & =  \nabla_{\theta}\log \rho(s_0)+\sum_{t=0}^H\nabla_{\theta}\log\left[ P(s_{t+1}|s_t,a_t)\right]\\
&\hspace{0.6in}+\sum_{t=0}^H \sum_{i=1}^n\nabla_{\theta}\log\left[\pi^i_{\theta^i}(a_t^i|o_t^i)\right]\\
&=\sum_{t=0}^H \sum_{i=1}^n\nabla_{\theta}\log\pi^i_{\theta^i}(a_t^i|o_t^i),
\end{aligned}
\end{equation}
and thus
\begin{equation}
\begin{aligned}
\nabla_{\theta^i}\log P(\tau|\pi_{\theta}) =\sum_{t=0}^H \nabla_{\theta^i}\log\pi^i_{\theta^i}(a_t^i|o_t^i).
\end{aligned}
\end{equation}
Putting the above equations together, we have the following
\begin{equation}
\begin{aligned}
\nabla_{\theta^i} J(\theta) &= \nabla_{\theta^i} \mathbb{E}[R(\tau)]\\
& = \nabla_{\theta^i}\int_{\tau}P(\tau|\pi_{\theta}) R(\tau)= \int_{\tau}\nabla_{\theta^i}P(\tau|\pi_{\theta}) R(\tau)\\
& = \int_{\tau}P(\tau|\pi_{\theta})\nabla_{\theta^i}\log P(\tau|\pi_{\theta}) R(\tau)\\
& = \mathbb{E}\left[\nabla_{\theta^i}\log P(\tau|\pi_{\theta}) R(\tau)\right]\\
& = \mathbb{E}\left[\sum_{t=0}^H \nabla_{\theta^i}\log\pi^i_{\theta^i}(a_t^i|o_t^i) R(\tau)\right].
\end{aligned}
\end{equation}
\end{appendices}
\end{document}